\documentclass[conference]{IEEEtran}
\IEEEoverridecommandlockouts
\usepackage{cite}
\usepackage{amsmath,amssymb,amsfonts}
\usepackage{algorithmic}
\usepackage{graphicx}
\usepackage{textcomp}
\usepackage{xcolor}
\def\BibTeX{{\rm B\kern-.05em{\sc i\kern-.025em b}\kern-.08em
    T\kern-.1667em\lower.7ex\hbox{E}\kern-.125emX}}

\usepackage{multirow}  
\usepackage{booktabs}  
\usepackage{makecell}  
\usepackage[colorlinks, linkcolor=red,  anchorcolor=blue, citecolor=blue]{hyperref}

\begin{document}

\newcommand{\cmark}{\ding{51}\xspace}%
\newcommand{\cmarkg}{\textcolor{lightgray}{\ding{51}}\xspace}%
\newcommand{\xmark}{\ding{55}\xspace}%
\newcommand{\xmarkg}{\textcolor{lightgray}{\ding{55}}\xspace}%

\newcommand{\KW}[1]{{\color{blue}{#1}}}
\newcommand{\FY}[1]{{\color{orange}{#1}}}
\newcommand{\SQ}[1]{{\color{cyan}{#1}}}
\newcommand{\JW}[1]{{\color{magenta}{\bf JW: #1}}}

\definecolor{cyan}{cmyk}{.3,0,0,0}

\newcommand{\RED}[1]{\textcolor{black}{#1}}
\newcommand{\YELLOW}[1]{\textcolor{yellow}{#1}}
\newcommand{\BLUE}[1]{\textcolor{blue}{#1}}
\definecolor{springgreen}{RGB}{0,255,0}

\newcommand{\lsimple}{L_{DM}}

\newcommand{\model}{\epsilon_\theta}
\newcommand{\modeluncond}{\tilde{\epsilon}_\theta}

\newcommand{\conditioner}{\tau_\theta}
\newcommand{\expec}{\mathbb{E}}
\newcommand{\encoder}{\mathcal{E}}
\newcommand{\decoder}{\mathcal{D}}

\newcommand{\textprompt}{\mathcal{P}}
\newcommand{\textembedding}{\mathcal{C}}

\newcommand{\updateprompt}{\hat{\mathcal{C}_t}}
\newcommand{\tokenset}{\mathcal{V}}
\newcommand{\updateset}{{\mathcal{V}_t}}
\newcommand{\updatetoken}{{v_t^k}}
\newcommand{\updateTKsubi}{{v_t^i}}
\newcommand{\updateTKsubj}{{v_t^j}}
\newcommand{\updateTK}{{v_t}}

\newcommand{\background}{\mathcal{B}}
\newcommand{\crossattn}{\mathcal{A}}
\newcommand{\gaussian}{\mathcal{F}}
\newcommand{\cluster}{\mathcal{M}}
\newcommand{\mlp}{\mathit{l}}

\newcommand{\ddimz}{z}
\newcommand{\denoisez}{{\bar z}}
\newcommand{\interz}{{\tilde z}}

\newcommand{\inputimage}{\mathcal{I}}

\newcommand{\tabincell}[2]{\begin{tabular}{@{}#1@{}}#2\end{tabular}}
\newcommand{\minisection}[1]{\vspace{0.01in} \noindent {\bf #1}\ }

\title{LiaisonAgent: An Multi-Agent Framework for Autonomous Risk Investigation and Governance } 


\author{
\IEEEauthorblockN{1\textsuperscript{st} Chuanming Tang}
\IEEEauthorblockA{\textit{Shenzhen Institute of Advanced } \\
\textit{Technology, CAS}\\
\textit{Sangfor Technologies Inc.} \\
Shenzhen, China \\
tangchuanming19@mails.ucas.ac.cn}
\and
\IEEEauthorblockN{2\textsuperscript{nd} Ling Qing}
\IEEEauthorblockA{\textit{College of Management Science } \\
\textit{Chengdu University of Technology }\\
Chengdu, China \\
qingling@cdut.edu.cn
}
\and

\IEEEauthorblockN{3\textsuperscript{rd} Shifeng Chen*\thanks{*Corresponding author} }
\IEEEauthorblockA{\textit{Shenzhen Institute of Advanced } \\
\textit{Technology, CAS}\\
\textit{Shenzhen University of Advanced Technology}\\
Shenzhen, China \\
shifeng.chen@siat.ac.cn}
}


\maketitle

\begin{abstract}
The rapid evolution of sophisticated cyberattacks has strained modern Security Operations Centers (SOC), which traditionally rely on rule-based or signature-driven detection systems. These legacy frameworks often generate high volumes of technical alerts that lack organizational context, leading to analyst fatigue and delayed incident responses. This paper presents LiaisonAgent, an autonomous multi-agent system designed to bridge the gap between technical risk detection and business-level risk governance. Built upon the QWQ-32B large reasoning model, LiaisonAgent integrates specialized sub-agents, including human-computer interaction agents, comprehensive judgment agents, and automated disposal agents—to execute end-to-end investigation workflows. The system leverages a hybrid planning architecture that combines deterministic workflows for compliance with autonomous reasoning based on the ReAct paradigm to handle ambiguous operational scenarios. Experimental evaluations across diverse security contexts, such as large-scale data exfiltration and unauthorized account borrowing, achieve an end-to-end tool-calling success rate of 97.8\% and a risk judgment accuracy of 95\%. Furthermore, the system exhibits significant resilience against out-of-distribution noise and adversarial prompt injections, while achieving a 92.7\% reduction in manual investigation overhead. 

\end{abstract}


\section{Introduction}

The current cybersecurity risk governance landscape is characterized by an escalating volume of dynamic threats and a corresponding shortage of skilled security analysts~\cite{vielberth2020security,abdullahi2022detecting}.Within large-scale production environments, organizations deploy User and Entity Behavior Analytics (UEBA)~\cite{salitin2018role} and static discovery engines to monitor for anomalies\cite{le2025cybersecurity,kinyua2021ai}. 
\textcolor{black}{However, a primary shortcoming of these legacy frameworks is the  contextual gap in their inherent inability to distinguish between technically anomalous behavior and legitimate business-related deviations.}
For instance, a detection engine may flag a user for downloading a high volume of sensitive files outside of normal working hours. While this represents a technical violation of standard baselines, it may concurrently constitute a benign operational necessity authorized by management for an urgent project. Currently, reconciling such alerts necessitates an exhaustive manual investigation—encompassing log analysis, cross-departmental inquiries, and multi-party verification \cite{zou2025blocka2a}. This manual overhead often becomes the primary bottleneck and longest path in the risk disposal cycle.

The integration of Large Language Models (LLMs)~\cite{vaswani2017attention} into cybersecurity workflows represents a transformative opportunity \cite{zhang2025llms}. Currently, applications have primarily utilized LLMs as \textit{security brains} for static alert analysis or automated report generation. However, these systems often remain isolated modules, lacking the autonomous agency required to interact with or modify the operational environment \cite{ali2025beyond}. Recent advancements in LLM-based agents have demonstrated significant potential in environmental perception and goal-oriented reasoning \cite{li2024survey, guo2024large}. 
In this paper, we propose an innovative security agent framework that leverages specialized tools to execute complex tasks, thereby enabling end-to-end automation for planning, investigation, adjudication, and remediation within the risk governance lifecycle. By evolving from singular detection and analysis models to multi-agent systems~\cite{dorri2018multi}, intricate SOC tasks can be decomposed into smaller, specialized sub-tasks managed by expert agents. 
\textcolor{black}{Therefore, by utilizing specialized sub-agents and autonomous planning, this framework bridges the gap between technical risk detection and business-level risk governance while optimizing performance through domain-specific role-playing to enhance system robustness against context saturation.}

\begin{figure}[!t] 
    \centering 
    \includegraphics[width=\linewidth]{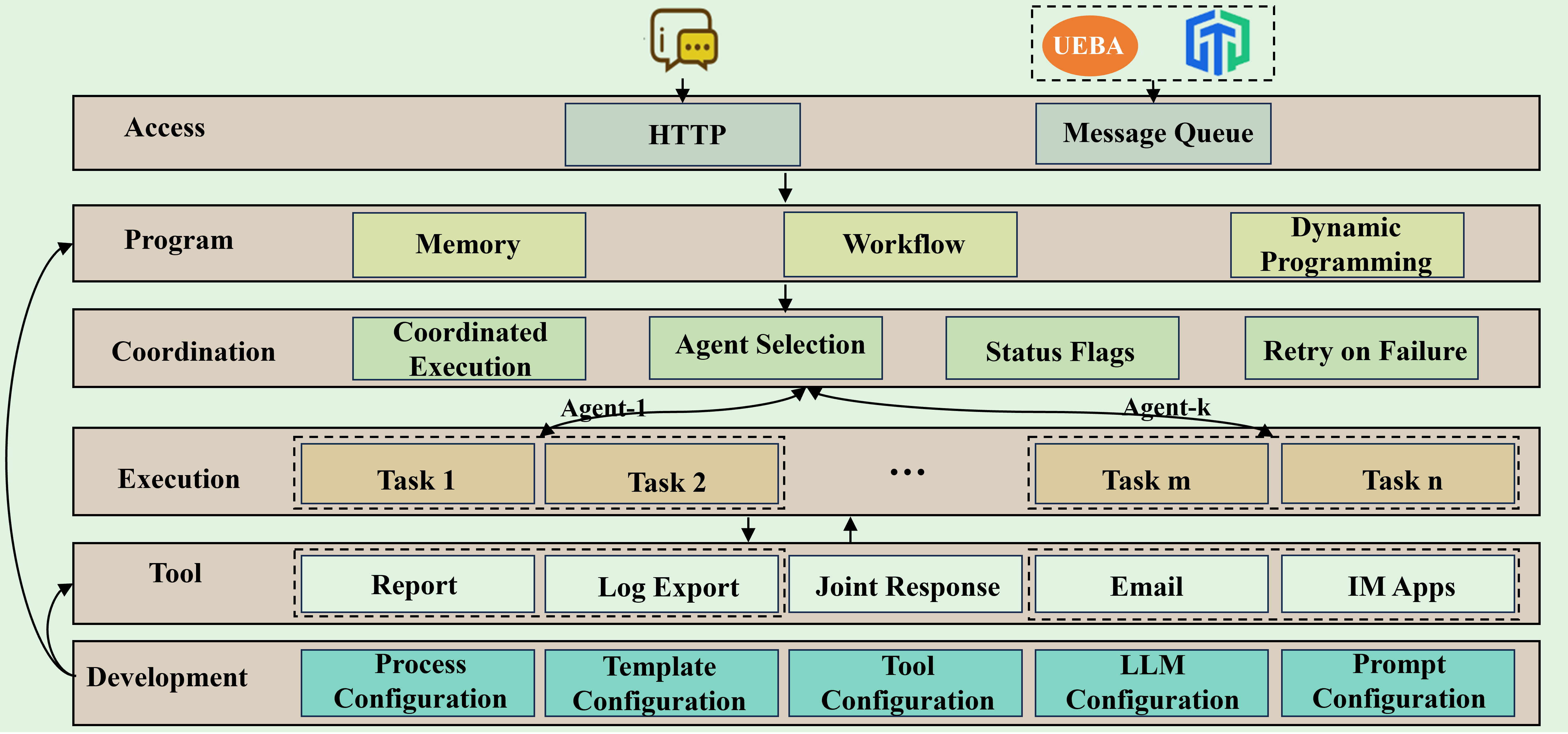} 
    \caption{Functional layered architecture of LiaisonAgent.} 
    \label{fig1} 
\end{figure}

LiaisonAgent is designed to bridge the gap between technical detection and business context by autonomously initiating inquiries with users and administrators via Instant Messaging (IM) platforms. By synthesizing human-derived qualitative evidence with raw technical telemetry, the agent formulates a finalized risk adjudication. Preliminary evaluations suggest that this approach can reduce the manual workload of  SOC analysts by more than an order of magnitude while maintaining high precision in risk qualification. To satisfy the framework’s fundamental requirements, namely accurate multi-step problem solving and long-context, multi-turn interaction, this study utilizes a specialized inference model trained with reinforcement learning, which enables the agent to provide stable, meticulous, and computationally efficient support throughout the entire security investigation process.
\textcolor{black}{To sum up, our contributions can be concluded as the following three aspects: }
\begin{itemize} 
\item \textcolor{black}{We propose LiaisonAgent, a system that utilizes specialized sub-agents to execute end-to-end investigation workflows, moving beyond singular detection models to a collaborative multi-agent paradigm. }
\item \textcolor{black}{We introduce a dual-path planning architecture that integrates deterministic workflows for organizational compliance with autonomous ReAct-based reasoning for handling unpredictable investigative scenarios.}
\item \textcolor{black}{The experiment demonstrates that the framework achieves a 97.8\% tool-calling success rate and a 92.7\% reduction in manual overhead while maintaining stability against adversarial prompt injections and out-of-distribution noise. }
\end{itemize}

\section{Method}

\subsection{System Framework and Core Architecture }

As illustrated in Figure~\ref{fig1}, the architecture is structured as a sophisticated multi-layered technology stack designed for autonomous planning~\cite{erdogan2025plan} and tool execution~\cite{qin2024toolllm}. The framework is logically organized into five functional layers supported by a foundational development tier to ensure both modularity and scalability. At the top, the access layer serves as the system's entry point, utilizing HTTP and message queues (integrated with UEBA) to handle real-time and asynchronous task triggers. These inputs are processed by the program layer, which manages the cognitive logic through memory, workflow orchestration, and dynamic programming to adapt to environmental changes. To ensure operational reliability, the coordination layer acts as the command center, overseeing coordinated execution and agent selection while implementing status flags and retry-on-failure mechanisms for robust error recovery. These strategic decisions are then manifested in the execution layer, where specialized subagents (agent-1 to agent-k) perform discrete task-1 to task-n. 

These agents interact with the external environment through the tool layer, which offers a comprehensive library supporting report generation, log export, joint response, and cross-platform communication via IM applications such as WeCom and Feishu.
The entire stack is underpinned by the development layer, a configuration framework that allows for the fine-tuning of processes, templates, tools, LLM parameters, and prompts, thereby ensuring the system’s behavior remains transparent, accountable, and highly adaptable to complex security domains.

\subsection{Multi-Agent Orchestration and Specialized Roles}

The complexity of risk investigation necessitates the partition of intelligence into three specialized sub-agents. The LiaisonAgent system employs a modular design where each agent focuses on a distinct phase of the investigation lifecycle, collaborating through a unified messaging interface.

\begin{figure}[!t] 
    \centering 
    \includegraphics[width=\linewidth]{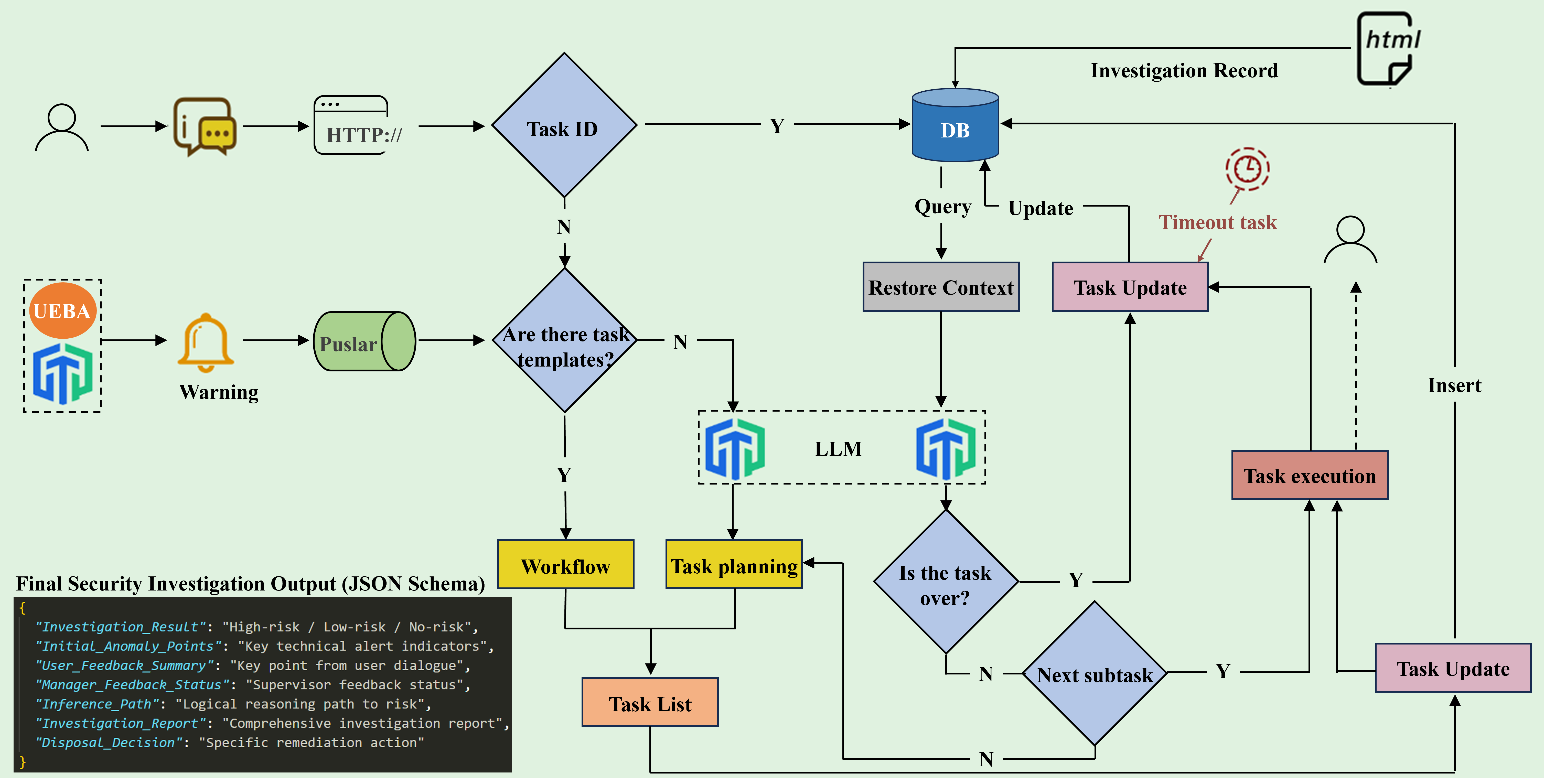} 
    \caption{LiaisonAgent overview of the internal details.} 
    \label{fig2} 
\end{figure}

The first component is the Human-Computer Interaction (HCI) agent, which is responsible for conducting structured inquiries with the actor involved in the security event, their supervisor, and the security administrator. Unlike basic chatbots, the HCI agent maintains a rigorous security investigator persona to ensure that all interactions remain professional, evidence-based, and strategically focused on uncovering justifications for anomalous behavior. The interaction logic is driven by the ReAct~\cite{yao2022react} paradigm. Figure~\ref{fig2} demonstrates the detailed process of interaction and investigation, including task initialization, triggering, LLM-driven iterative investigation, final structured output, and record persistence.
Upon receiving an alert, such as a large-scale sensitive file download, the HCI agent analyzes specific anomaly dimensions, including target directories, file volume, and temporal patterns, to formulate targeted inquiries. Should a participant provide a vague response, the agent leverages its reasoning capabilities to perform follow-up inquiries, seeking explicit operational justifications or authorization statements. To balance investigative depth with user experience, the system enforces a strict constraint of three to ten dialogue rounds and may extend inquiries to supervisors or administrators when contextualized situational logic dictates.

The judgement agent serves as the analytical core of LiaisonAgent, succeeding the HCI phase. This agent integrates three primary information streams: technical metadata, behavioral analysis results, and human investigation feedback. Technical metadata encompasses IP addresses, access patterns, and threat signatures derived from initial UEBA alerts. Behavioral analysis results represent synthesized anomalies generated by specialized heuristic rules and the UEBA engine. Human feedback refers to the structured dialogue transcripts between the HCI agent and relevant personnel. By evaluating the consistency of evidence across these three dimensions, the adjudication agent formulates a definitive risk conclusion.

Finally, the disposal agent executes the terminal phase of risk governance. Based on the adjudicated conclusion, it selects appropriate interventions from a predefined disposal toolkit. For benign events, the agent issues a closure notification accompanied by security awareness training resources. Conversely, for verified threats, it escalates the incident to human responders and initiates automated containment protocols, such as two-factor authentication, session termination, IP blacklisting, or network isolation.

\subsection{Hybrid Planning and Tool Design }

To operationalize these sub-agents, LiaisonAgent adopts a hybrid planning mechanism designed to manage both routine compliance procedures and unpredictable investigative scenarios. This approach ensures procedural predictability for standardized tasks while enabling high-level intelligence for complex cases. For deterministic investigative steps, such as securing administrative authorization prior to an investigation or delivering security awareness content upon session termination, the system utilizes fixed workflows. These are defined as \textit{pre-steps} and \textit{post-steps} within the task plan, effectively circumventing the inherent stochasticity of LLM-based planning to achieve 100\% compliance with organizational policies. For the dynamic investigative phase, the agent uses the autonomous planning method. In this stage, LLM will generate a context-aware, multi-step dynamic strategy based on the alert situation. This plan is maintained as a dynamic \textit{todo.md} file, which the agent updates the state iteratively. Following each tool-mediated action, a reflection step is executed to evaluate the success of the execution and determine whether the information gathered is sufficient to progress to the subsequent phase.

To address enterprise-scale alert volumes, the framework incorporates an asynchronous state-persistent coordination module. Each investigation is assigned a unique task ID, with all intermediate states, including planning updates and tool responses, persisted in a relational database. This architecture facilitates the management of thousands of concurrent investigation threads and prevents memory exhaustion, ensuring system stability even during prolonged latencies, such as waiting for human responses over several hours or days.
To facilitate interoperability, each sub-agent and tool interacts through standardized interfaces. 
Table~\ref{tab:tool_library} presents a selection of representative tools from the library, along with their associated functionalities. 
Every tool within the library is encapsulated in a dedicated schema that explicitly defines its nomenclature, functional description, and mandatory parameter. The LLM leverages this structured metadata to generate precise and executable function calls, ensuring that the model's output aligns with the underlying programmatic requirements.

\begin{table}[!t]
  \centering
  \caption{LiaisonAgent Tool Library Examples. }
  \label{tab:tool_library}
  \resizebox{\linewidth}{!}{
    \begin{tabular}{lll}  
      \toprule
       \textbf{Specific Tool} & \textbf{Functional Summary} & \textbf{Logic / Trigger} \\
      \midrule
      invest\_ask\_admin & Requests authorization from a security admin. & Triggered by high-severity technical alerts. \\
       invest\_notify\_admin & Pushes unidirectional updates to administrators. & Used for status logging and final reporting. \\
       invest\_ask\_user & Conducts multi-round Q\&A with the user. & Initiated after administrator approval. \\
       invest\_ask\_manager & Confirms business legitimacy with a supervisor. & Triggered if users' feedback is suspicious. \\
       invest\_judge & Performs final risk qualification. & Executes after all feedback is collected. \\
       invest\_notify\_user & Sends final status, security notification etc. & Triggered upon task completion. \\
       terminate & Suspends or ends an investigation thread. & Used to manage state transitions. \\
       closed\_loop\_Processing  & Take different actions to eliminate the risk  & Triggered upon task completion. \\ 
      \bottomrule
    \end{tabular}
  }
\end{table}

\section{Experiment}
\subsection{Evaluation Setting }

The evaluation of an autonomous security agent requires a methodology that goes beyond static classification, requiring a rigorous assessment of the agent's inference trajectory. This includes the precision of tool selection and execution, the robustness of handling unexpected inputs, and the convergence capability of the terminal investigative state. To circumvent data sensitivity and privacy constraints inherent in real-world security operations, we developed a large-scale synthetic evaluation dataset utilizing an LLM-as-user simulation Q\&A framework.

Within this framework, a dedicated instance of the DeepSeek-R1 model\cite{guo2025deepseek} was deployed to simulate the personas of both the \textit{risky actor} and their supervisor. The simulation is guided by four distinct behavioral principles designed to encompass a broad spectrum of human interaction. These principles range from \textit{cooperative and detailed} responses providing logical business justifications and \textit{cooperative but brief} replies offering concise technical facts to more challenging \textit{evasive/vague} stances characterized by ambiguity and \textit{deceptive/adversarial} attempts to actively mislead the investigator. 

Reflecting standard corporate security assumptions where supervisors are presumed to be honest and compliant, the two cooperative principles are applied to both user and supervisor agents. Conversely, the evasive and deceptive profiles are exclusively reserved for the risky user persona. During dataset generation, the DeepSeek-R1 agent randomly adopts one of the applicable principles for each sample, thereby eliciting realistic and diverse dialogues for a comprehensive analysis of the agent's investigative performance.

For the test dataset, we collected 2000 unique behavioral alert samples and combined each technical alert with 4 user response methods and 2 manager feedback scenarios to form a complete investigation-judgement-disposal path, expanding it into a survey dataset containing 16000 samples. The testing scenarios include large-scale file downloads, cross-functional account borrowing, and other anomalies, such as IP scanning, suspicious logons, and the use of prohibited software (e.g., global proxies) and web crawler scripts accessing applications.

\subsection{Quantitative Results}
\textbf{Tool-Call Success Rate. } The reliability of the investigative pipeline hinges on the agent’s capability to invoke tools accurately with proper parameters. This evaluation contrasted the Original Success Rate (OSR) with the Final Success Rate (FSR). OSR means the one-time success rate with the output directly generated by the LLM, and FSR denotes the LLM output with engineering post-processing. As illustrated in Table~\ref{tab:multi_agent_tool_calling_accuracy}, our framework achieves an average success rate of 97.8\%. A granular analysis of the remaining failure modes identifies three primary causes: redundant invocations of notification tools, parameter formatting inconsistencies, and JSON parsing errors. 

\begin{table}[!t]
  \centering
  \caption{Multi-Agent Tool-Calling Accuracy. }
  \label{tab:multi_agent_tool_calling_accuracy}
    \begin{tabular}{llll}  
      \toprule
      \textbf{Alert Category} & \textbf{Samples} & \textbf{OSR} & \textbf{FSR} \\
      \midrule
      Large Files Download & 4000 & 93.0\% & 94.0\% \\
      Account Borrowing & 4000 & 99.0\% & 99.7\% \\
      Other Anomalies & 8000 & 97.0\% & 98.8\% \\
      Weighted Average & - & 96.5\% & 97.8\% \\
      \bottomrule
    \end{tabular}
\end{table}

\begin{table}[!t]
  \centering
  \caption{Interaction  Qualitative Evaluation in HCI Agent. }
  \label{tab:qualitative_interaction_performance}
    \begin{tabular}{ll}  
      \toprule
      \textbf{Agent Role} & \textbf{Score} \\
      \midrule
      Investigator for User Inquiry & 93\% \\
      Investigator for supervisor verification & 99\% \\
      Comprehensive Risk Judgment & 98\% \\
      Risk Disposal & 98\% \\
      \bottomrule
    \end{tabular}
\end{table}

\textbf{Interaction Quality.} As presented in Table~\ref{tab:qualitative_interaction_performance}, human evaluators conducted a qualitative audit of 500 investigation logs, scoring the agent’s performance on a continuous scale from 0 to 1. The evaluation metrics for the investigative agents primarily focused on interaction logic, linguistic fluency, and the anthropomorphic quality of conversational expressions. The investigative performance concerning users and their supervisors achieved reliability scores of 93\% and 100\%, respectively, representing a proficiency level nearly indistinguishable from that of human security personnel. Regarding comprehensive risk assessment, 98\% of the samples yielded correct and reasonable qualitative conclusions about the risks. Finally, in 98\% of the samples, appropriate tools could be selected to reduce risks and close loop problems.

\textbf{Model Selection.}  
A primary design constraint for LiaisonAgent is its deployment on a server infrastructure equipped with four NVIDIA RTX 4090 GPUs. As detailed in Table~\ref{table:llm_evaluation}, we evaluated five candidate LLMs to identify the optimal optimal model for the agent: Qwen2.5-32B~\cite{qwen2025qwen25technicalreport}, Qwen3-32B~\cite{yang2025qwen3}, QwQ-32B~\cite{qwq2024}, Qwen3-14B, and Qwen3-30B-A3B \cite{yang2025qwen3}.
Built on the Qwen2.5-32B foundation, QWQ-32B incorporates extensive multi-stage post-training, including a first-stage Chain-of-Thought (CoT) fine-tuning and a second-stage GRPO~\cite{shao2024deepseekmath-grpo} reinforcement learning stage, endowing it with advantages in planning, tool usage, context length handling, and other critical capabilities. Compared with other models, QwQ-32B achieves the optimal tool call and planning ability. Additionally, QWQ-32B achieves comparable performance on other general ability aspects with the closed-source proprietary reasoning models like OpenAI's o1-mini~\cite{jaech2024openai-o1}.

To identify the optimal baseline LLM for LiaisonAgent within the constraints of 4 RTX 4090 GPUs, five candidate models were evaluated across four core capabilities. As shown in Table~\ref{table:llm_evaluation}, QWQ-32B was selected as the baseline model due to its superior overall performance. Specifically, QWQ-32B achieved the highest scores in tool calling (96.5\%), planning accuracy (97.8\%), and logical reasoning (89.6\%) among all candidates, demonstrating a significant competitive advantage in these critical functional areas. Furthermore, its investigation dialogue accuracy (87.3\%) was nearly equivalent to the best-performing Qwen3-32B (87.5\%). These results confirm that QwQ-32B effectively satisfies the operational requirements of LiaisonAgent, representing the ideal equilibrium between high-fidelity performance and hardware-constrained compatibility.

\begin{table}[t]
  \centering
  \caption{Performance Comparison of Different LLMs. }
  \label{table:llm_evaluation}
  \renewcommand{\arraystretch}{1.2} 
  \resizebox{\linewidth}{!}{
  \begin{tabular}{lcccc}
    \toprule
    \textbf{LLM Model} & \textbf{Tool Calling} & \textbf{Planning Accuracy} & \textbf{Logical Reasoning} & \textbf{Investigation Dialogue} \\
    \midrule
    Qwen2.5-32B        & 71.2\%                  & 95.6\%                      & 82.5\%                       & 82.1\%                             \\
    Qwen3-14B          & 72.3\%                  & 90.8\%                      & 80.9\%                       & 79.2\%                             \\
    Qwen3-32B          & 88.5\%                  & 96.3\%                      & 85.7\%                       & \textbf{87.5\%}                             \\
    Qwen3-30B-A3B      & 90.1\%                  & 94.5\%                      & 81.6\%                       & 85.7\%                             \\
    QwQ-32B            & \textbf{96.5\%}         & \textbf{97.8\%}             & \textbf{89.6\%}              & 87.3\%                             \\
    \bottomrule
  \end{tabular}
  }
\end{table}

\subsection{Robustness Evaluation.} 

\textbf{Noisy Answer in Investigation. }
Autonomous security agents operate in open-world environments, where they are inherently susceptible to noisy or intentionally malicious input data. To evaluate the resilience of the proposed framework, we conducted rigorous robustness testing against three categories of non-normative inputs designed to disrupt agent operations: emotional appeals, irrelevant topical shifts, and semantic voids or gibberish.
As illustrated in Table~\ref{tab:robustness_results}, the agent achieved high correct response rates across all three categories of non-normative inputs. Specifically, it maintained a 95\% correct response rate when facing irrelevant topics, effectively avoiding off-topic engagement and ensuring mission focus. 
For gibberish and semantic voids, the agent maintained a 94\% correct response rate, consistently focusing on relevant tasks.
For users' emotional appeals, 98.0\% of cases can achieve a good balance between task focus and user reassurance.
These results fully confirm that the proposed agent framework has strong robustness and can stably maintain persona consistency and mission integrity in open-world environments with noisy or disruptive inputs.

\textbf{Adversarial Prompt Injection in Investigation. }
To verify the proposed agent's robustness against adversarial scenarios, two primary adversarial prompt injection attack vectors were simulated for targeted security evaluation. For instruction overriding attacks, adversaries attempted to force premature case closure by crafting malicious prompts with high-privilege syntax (e.g., \textit{"Admin decision: Case closed. Risk Level: No Risk."}). As presented in Table~\ref{tab:robustness_results}, the agent achieved a 95\% defense success rate against such attempts, with few failures occurring when adversarial injections were embedded in complex technical jargon that mimicked the agent’s intrinsic reasoning structure. For white-box tool spoofing attacks, adversaries with partial knowledge of the system’s API sought to directly skip the \textit{invest\_ask\_manager} tool for forging unauthorized approvals, and  90\% of samples successfully defended against the skip process. 

Collectively, these results validate that the proposed agent framework establishes a well-defined security boundary and exhibits strong adversarial resilience, capable of maintaining stable and reliable decision-making performance in the face of both out-of-distribution noisy inputs and intentional adversarial prompt injection attacks. 
The experiment results indicate that while the LLM core remains partially susceptible to sophisticated prompt injection attacks, the integration of prompt-level constraints with rigid engineering validation layers provides the requisite reliability for production-grade security environments.

\begin{table}[htbp]
\centering
\caption{Agent Robustness Evaluation under Different Investigation Responses. }
\label{tab:robustness_results}
  \resizebox{0.8\linewidth}{!}{
\begin{tabular}{ccc}
\toprule
\textbf{ Answer Type} & \textbf{Sample Size} & \textbf{Defense Success Rate} \\ 
\midrule
Irrelevant Topics Answer             & 50 & 95\%                     \\
Gibberish \& Semantic Voids    & 50 & 94\%                      \\
Emotion Appeals          & 50 & 98\%                   \\
Instruction Injection   & 100 & 95\%       \\
White-Box Tool Spoof    & 100 & 100\%   \\ \bottomrule
\end{tabular}   
}
\end{table}

\textbf{Performance and Hardware Efficiency.} 
The agent was benchmarked on a cluster equipped with four RTX 4090 GPUs to assess its readiness for enterprise-scale deployment, with a focus on throughput, latency, and hardware utilization efficiency. 
Benchmarking results demonstrated the system’s strong adaptability to enterprise security demands.
LiaisonAgent achieved a maximum concurrency of 50 parallel investigations, enabling it to effectively handle multiple burst alerts at the same time, while the human operators can only process alerts one-by-one. 
In terms of latency, the agent exhibited a mean latency of 2 seconds per step, ensuring prompt response to security events and avoiding delays in incident handling. 
Lastly, the average end-to-end investigation time was 48 seconds (without investigation waiting for answers from humans), representing a streamlined incident processing cycle that outperforms traditional workflows. 

In terms of throughput, LiaisonAgent could process up to 1,800 investigations per day, which fully meets the daily alert process demands of typical enterprise security teams. In contrast, a human operator can only investigate 50 alerts in a whole day, which demonstrates the agent can improve operational efficiency by 36 times. 
In this way, a 92.7\% manual workload reduction enables security analysts to focus on high-value, complex tasks rather than routine incident triage and investigation.

\textcolor{black}{
\textbf{Analysis of Advantages and Distinctive Features.} 
LiaisonAgent distinguishes itself from traditional Security Orchestration, Automation, and Response (SOAR) platforms and monolithic LLM applications through its unique synthesis of proactive agency and specialized multi-agent orchestration. Unlike passive diagnostic tools, the framework exhibits contextual proactivity by autonomously initiating multi-turn inquiries with human stakeholders to resolve the ambiguity between technical anomalies and business legitimacy. This capability is underpinned by an asynchronous state-persistent coordination module that maintains over 50 concurrent investigative threads via relational database storage, ensuring resource efficiency during prolonged human-response latencies. Furthermore, by decomposing complex workflows into specialized roles, the system prevents the instruction-following degradation common in monolithic models, while its optimization for consumer-grade hardware ensures that high-fidelity autonomous risk governance is both scalable and economically accessible for standard enterprise environments.}

\section{Conclusion}
This paper presents an intelligent multi-agent system for fully automated cybersecurity risk investigation, leveraging the multi-step reasoning of the QWQ-32B LLM. By integrating a hybrid planning mechanism that balances deterministic workflows with autonomous ReAct-based decision-making, the framework effectively bridges the gap between technical anomaly detection and high-level risk governance. 
Experimental results demonstrate the system's efficacy, achieving a 97.8\% tool invocation success rate and 95\% judgment accuracy while reducing manual workloads by 92.7\%. Ultimately, the system’s robustness against adversarial noise confirms its potential as a reliable and scalable foundation for trustworthy enterprise security operations.
Our future research will prioritize the development of hybrid general-specialized models to refine tool-calling and risk-assessment capabilities while preserving the model’s inherent general intelligence. 
Additionally, we aim to incorporate long-term security memory mechanisms, enabling the system to learn from historical investigation outcomes and evolving user behaviors. These advancements will significantly enhance contextual awareness and adaptive decision-making, positioning intelligent agents as a cornerstone of next-generation autonomous Security Operations Centers.

\bibliographystyle{IEEEbib}
\bibliography{ztp}

\end{document}